\documentclass{iaus}
\usepackage{graphicx}

\title[Few dwarfs in Fornax] 
{A flat faint end of the Fornax cluster galaxy luminosity function}

\author[S. Mieske et al.]   
{S. Mieske$^1$ 
,
 M. Hilker$^1$, L. Infante$^2$ \and C. Mendes de Oliveira$^3$}

\affiliation{$^1$European Southern Observatory, Karl-Schwarzschild-Str.2, 85748 Garching b. M\"unchen, Germany\break email: smieske@eso.org,mhilker@eso.org\\[\affilskip]
$^2$Departamento de
     Astronom\'{\i}a y Astrof\'{\i}sica, Pontificia
     Universidad Cat\'olica de Chile, Casilla 306, Santiago 22, Chile \break email: linfante@astro.puc.cl \\[\affilskip]$^3$Instituto de Astronomia, Geof\'isica, e Ci\^encias Atmosf\'ericas, Departamento de Astronomia, Universidade de S\~ao Paulo, Rua do Mat\'ao 1226, Cidade Universit\~aria, 05508-900 S\~ao Paulo, SP, Brazil \break email: oliveira@astro.iag.usp.br}

\pubyear{2007}
\volume{244}  
\pagerange{119--123}
\date{?? and in revised form ??}
\jname{IAU Symposium 244: Dark Galaxies and Lost Baryons}
\editors{J.I. Davies, M.D. Disney}
\begin{document}

\maketitle

\begin{abstract}
  {We analyse the photometric properties of the early-type Fornax
    cluster dwarf galaxy population ($M_V>-17$ mag), based on a wide
    field imaging study of the central cluster area in $V$ and $I$
    band-passes with IMACS/Magellan at Las Campanas Observatory.  We
    create a fiducial sample of $\sim$ 100 Fornax cluster dwarf
    ellipticals (dEs) with $-16.6<M_V<-8.8$ mag in the following three
    steps: (1) To verify cluster membership, we measured $I$-band
    surface brightness fluctuations (SBF) distances to candidate dEs
    known from previous surveys; (2) We re-assessed morphological
    classifications for those candidate dEs that are too faint for SBF
    detection; and (3) We searched for new candidate dEs in the
    size-luminosity regime close to the resolution limit of previous
    surveys.  The resulting fiducial dE sample follows a well-defined
    surface brightness - magnitude relation, showing that Fornax dEs
    are about 40\% larger than Local Group dEs. The sample also
    defines a colour-magnitude relation similar to that of Local Group
    dEs. The early-type dwarf galaxy luminosity function in Fornax has
    a very flat faint end slope $\alpha \simeq -1.1 \pm 0.1$. We
    compare the number of dwarfs per unit mass with those in other
    environments and find that the Fornax cluster fits well into a
    general trend of a lack of high-mass dwarfs in more massive
    environments.}  
\end{abstract}

\firstsection 
\section{Introduction}
One of the most important quantities for characterising a galaxy
population is the galaxy luminosity function (GLF). Its logarithmic
faint-end slope $\alpha$ is a very useful quantity to be contrasted
with the expected slope for the mass spectrum of cosmological
dark-matter halos (e.g. Jenkins {\it et~al.}~\cite{Jenkin01}, Moore {\it et
al.}~\cite{Moore99}). Generally, the value of $\alpha$ derived in
various environments including the Local Group is much shallower than
the expected slope of dark matter halos (see for example Grebel {\it et
al.}~\cite{Grebel03}; Trentham \& Tully~\cite{Trenth02}; Trentham {\it et
al.}~\cite{Trenth05}; Andreon {\it et~al.}~\cite{Andreo06}; Tanaka {\it et
al.}~\cite{Tanaka04}; and Infante {\it et~al.}~\cite{Infant02} and
references therein). This discrepancy is also known as the
``substructure problem'' of present-day cosmology.

In this contribution we focus on the GLF in the nearby Fornax cluster
of galaxies. Up to now, investigations of the Fornax GLF in the low
luminosity regime ($M_{\rm V}>-14$ mag) have been restricted to
morphological cluster-membership assignment, given the lacking depth
of spectroscopic surveys. The important restriction of the
morphological assessment is the uncertainty in estimating the amount
of contamination by background galaxies (e.g. Trentham \&
Tully~\cite{Trenth02}). This can lead to different authors deriving
very different slopes for the same cluster: Ferguson \&
Sandage~(\cite{Fergus88}) obtain $\alpha = -1.08 \pm 0.09$ for the
dwarf GLF in Fornax; Kambas {\it et~al.}~(\cite{Kambas00}) suggest a much
steeper slope $\alpha \simeq -2.0$, based on poorer resolution data of
$2.3''$ without colour information (see also the discussion in Hilker
{\it et~al.}~\cite{Hilker03}). Such differences in $\alpha$ stress the
need for high-resolution imaging and an extension of the limiting
magnitude for direct cluster membership determination.

In this contribution (see also Mieske {\it et~al.}~\cite{Mieske07}), we
investigate the photometric properties of the dwarf galaxy population
in the central 1.5 degree of the Fornax cluster ($M_{\rm V}>-17$
mag). For this we have used the wide-field imaging instrument IMACS
mounted at the 6.5m Magellan telescopes at LCO, yielding a $0.2''$
pixel scale. This work follows up a previous study with the 2.5m du
Pont telescope and the WFCCD camera (Hilker {\it et~al.}~\cite{Hilker03},
based on $0.8''$ pixels. To overcome the difficulties of
morphological cluster membership assignment, we adopt the following
three steps to define a fiducial sample of Fornax cluster dwarf
ellipticals.

1. Confirm cluster membership of candidate dEs via measuring surface
brightness fluctuations (SBF, Tonry \& Schneider~\cite{Tonry88}), and
hence their distance.

2. Re-assess the morphological classification of candidate dEs too
faint for SBF.

3. Search for new candidate dEs close to the resolution limit of
previous surveys.

\begin{figure}
\centering
\resizebox{4.1cm}{!}{\includegraphics{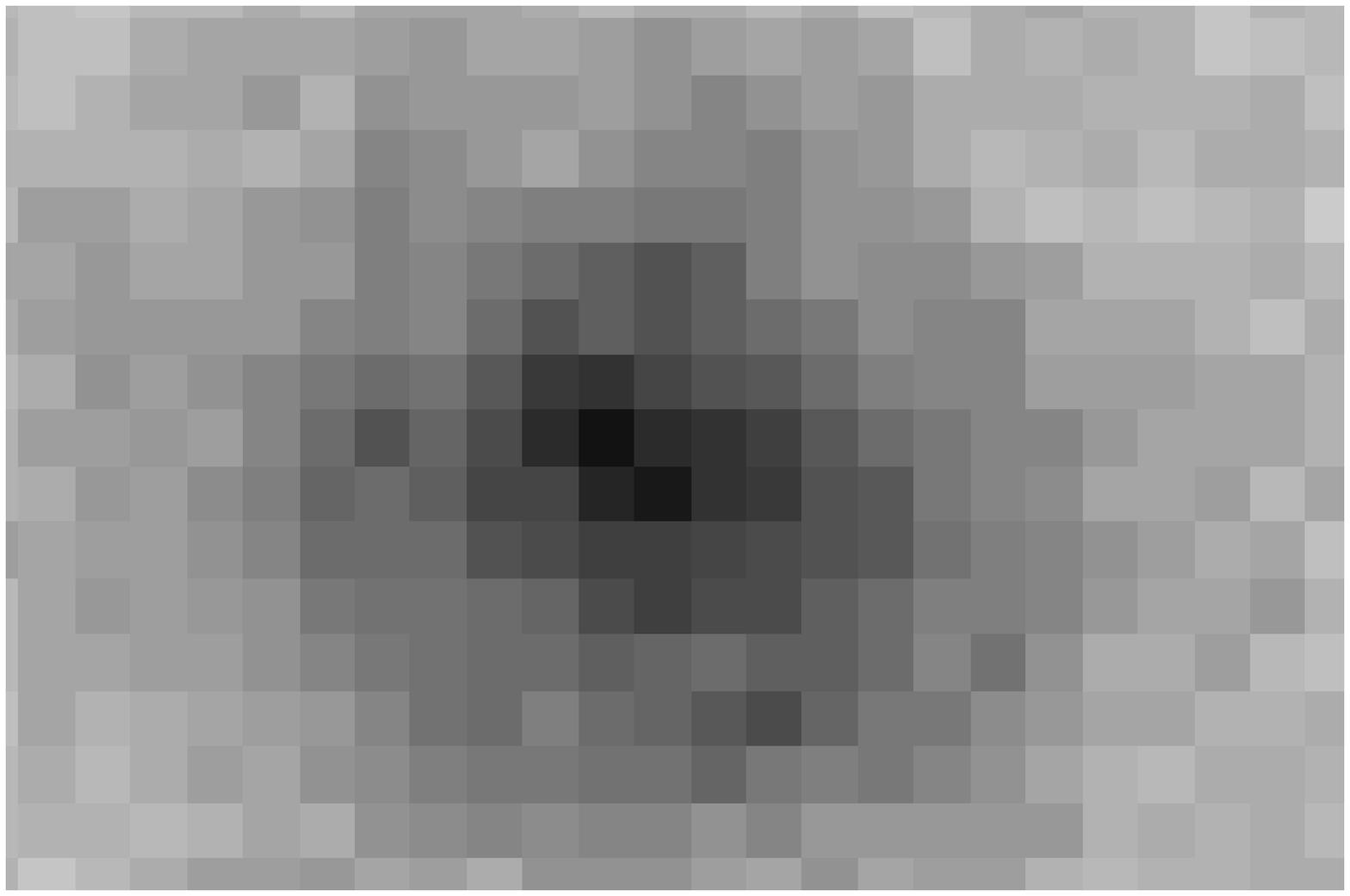} }
\resizebox{4.1cm}{!}{\includegraphics{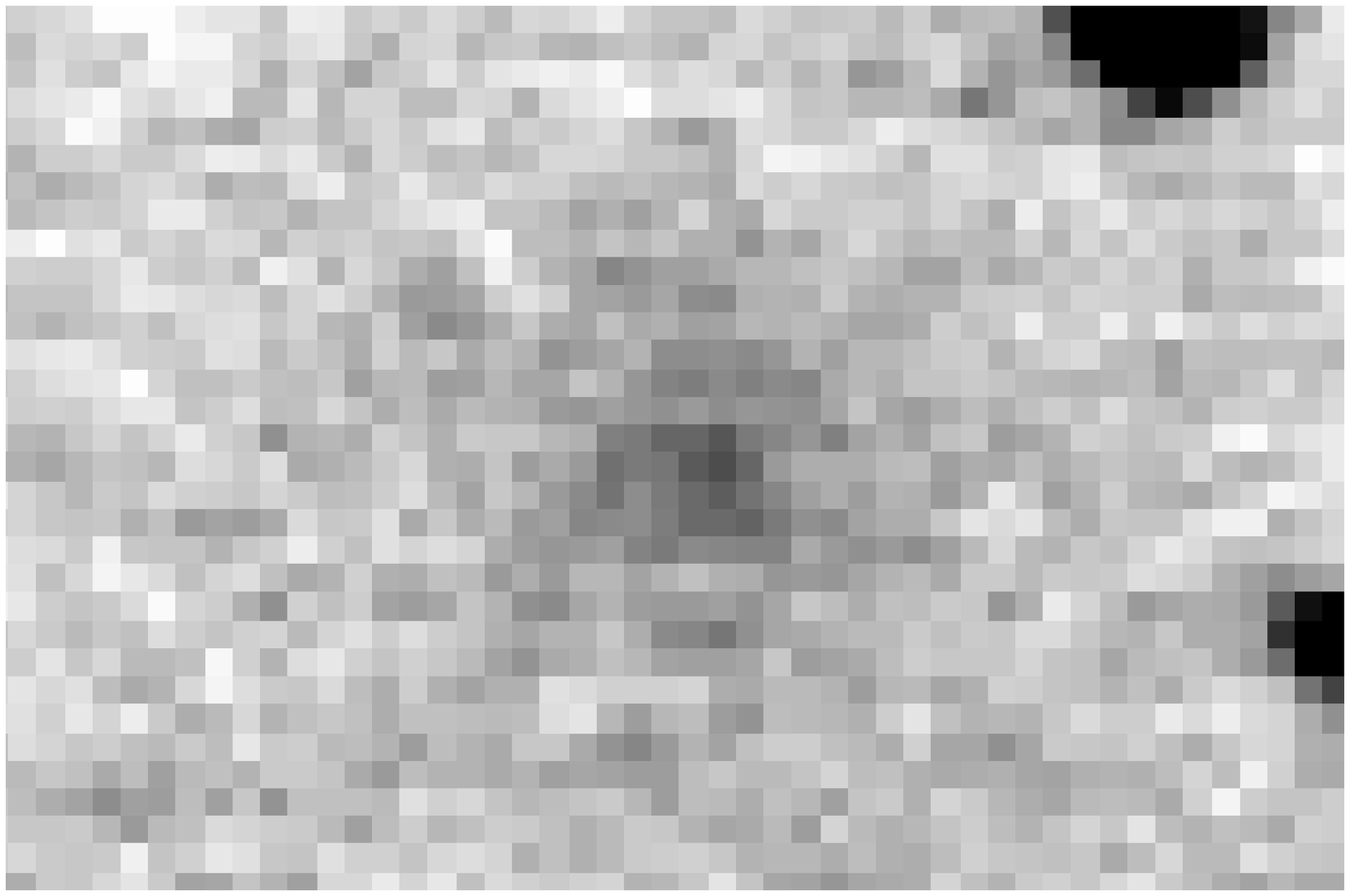}}
\resizebox{4.1cm}{!}{\includegraphics{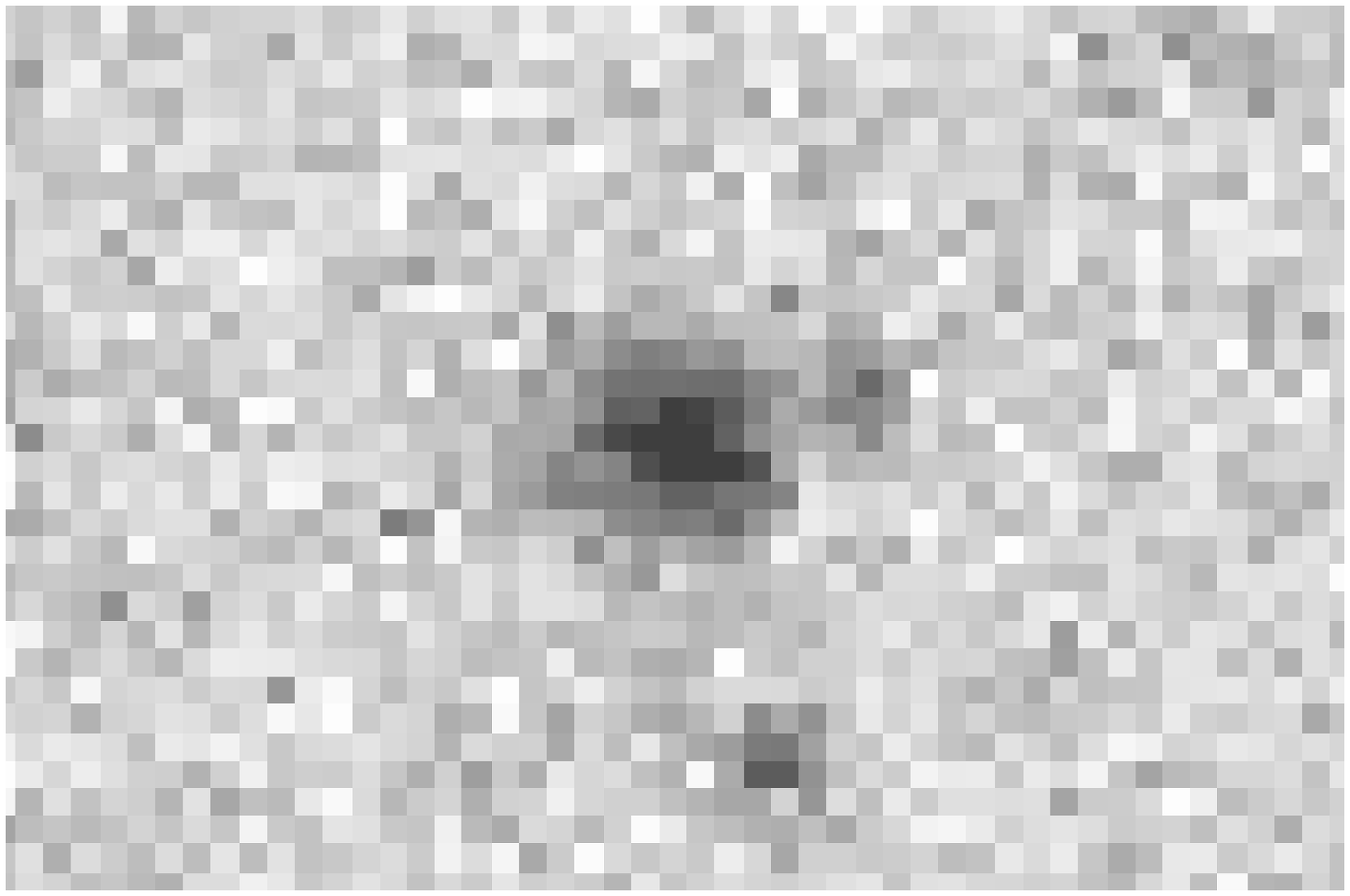} }\vspace{0.3cm}\\
\resizebox{4.1cm}{!}{\includegraphics{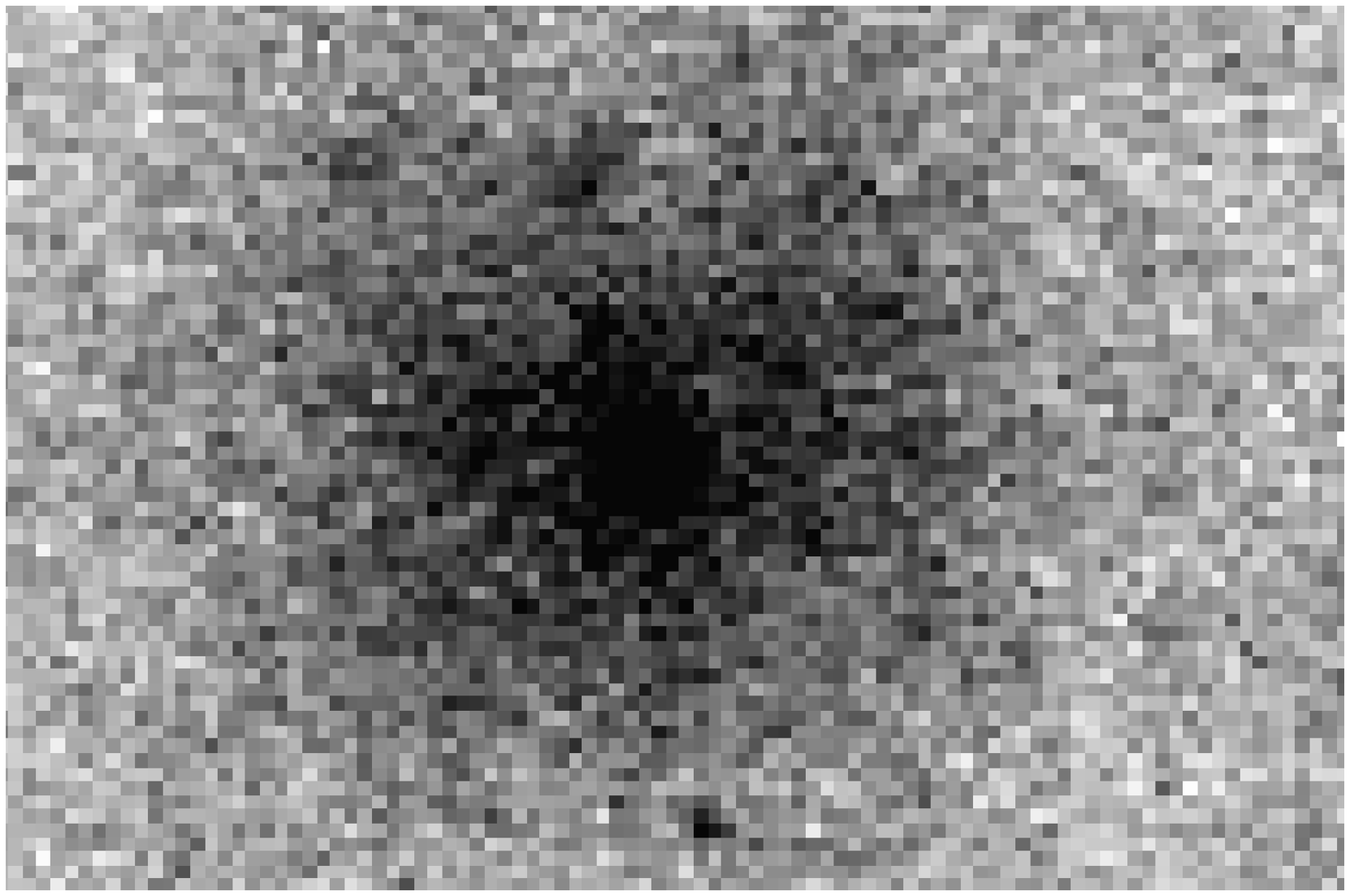} }
\resizebox{4.1cm}{!}{\includegraphics{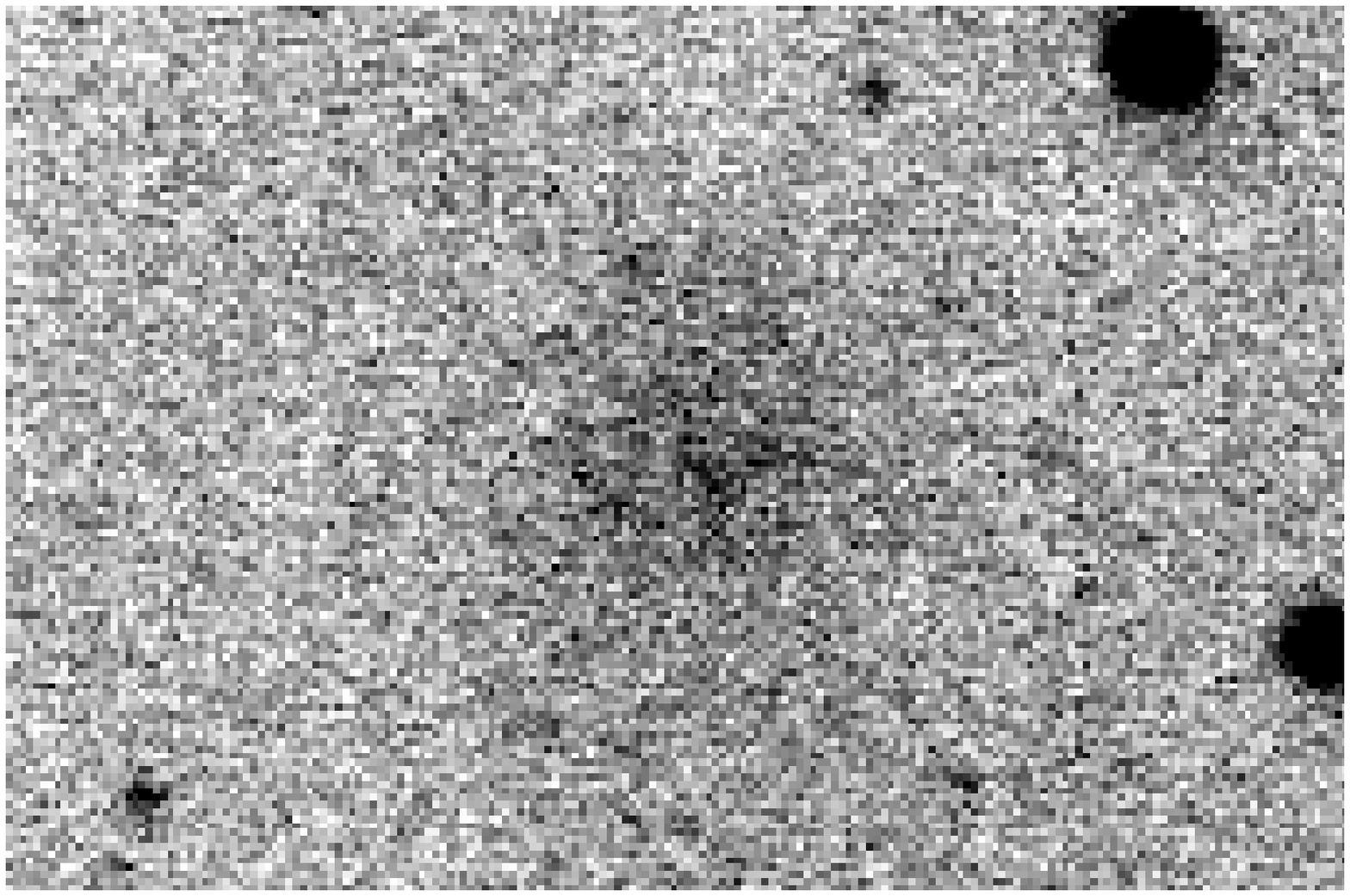}}
\resizebox{4.1cm}{!}{\includegraphics{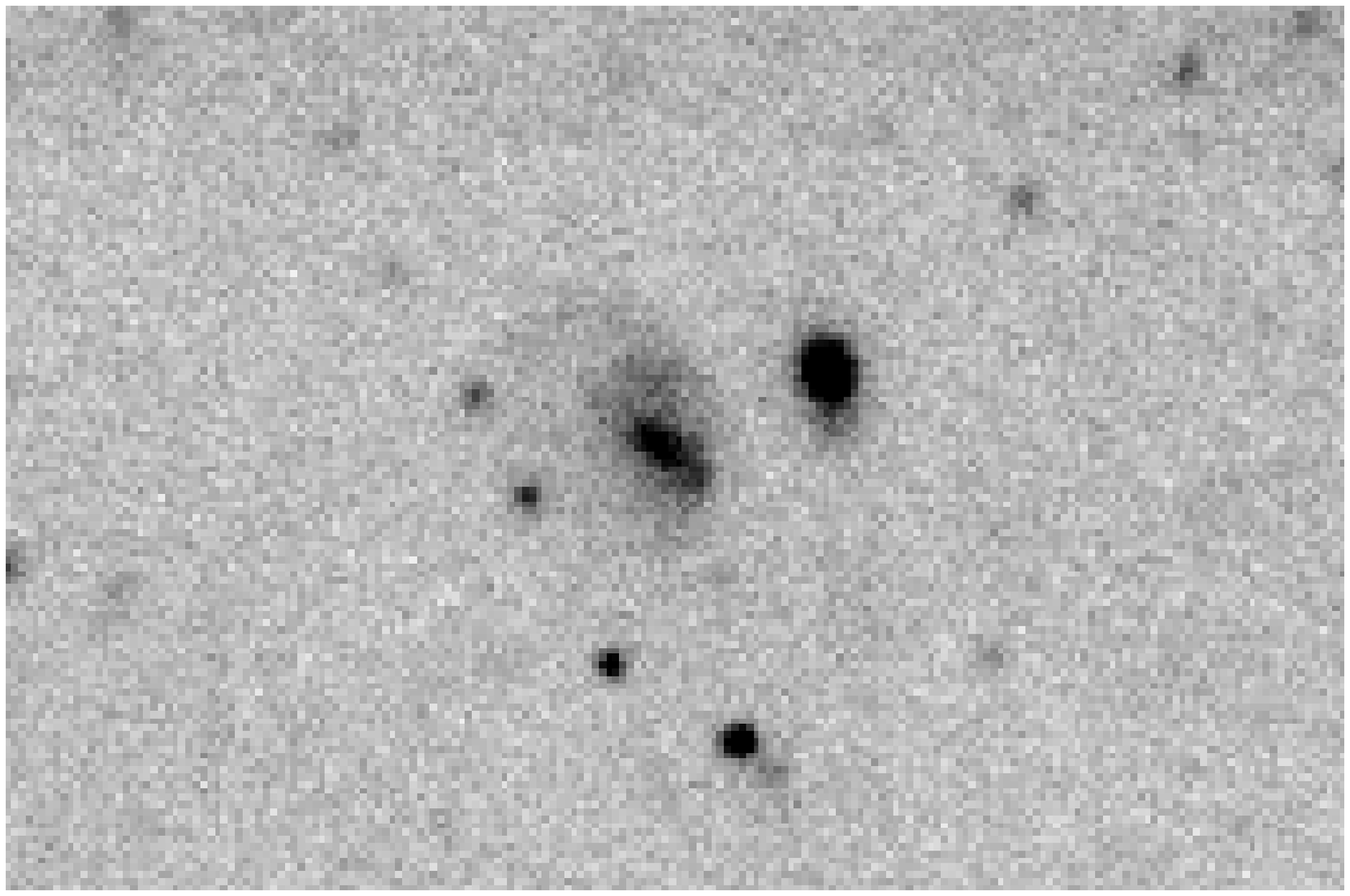} }\\

\caption[]{{\bf Top:} Example thumbnails of three dwarf galaxy candidates from the sample of Hilker {\it et~al.} (2003). {\bf Bottom:} The same three galaxies,
imaged with the higher resolution IMACS data (Mieske {\it et~al.} 2007). Galaxies are from left to right: FCC 191, which is confirmed via SBF measurement with the higher resolution IMACS data; WFLSB2-1, whose morphological assessment as probable cluster member from Hilker {\it et~al.} (2003) is confirmed with the IMACS data; WFLSB6-3, whose morphological assessment as probable cluster member from Hilker {\it et~al.} (2003) is changed to probable background.}
\label{fig1}
\end{figure}

The details of these procedures are described in Mieske {\it et
al.}~(\cite{Mieske07}). Fig.~\ref{fig1} gives example thumbnails of dwarf
galaxy candidates, showing images from both the lower resolution WFCCD data
and the higher resolution IMACS data.

\begin{figure}
\centering
\resizebox{6.5cm}{!}{\includegraphics{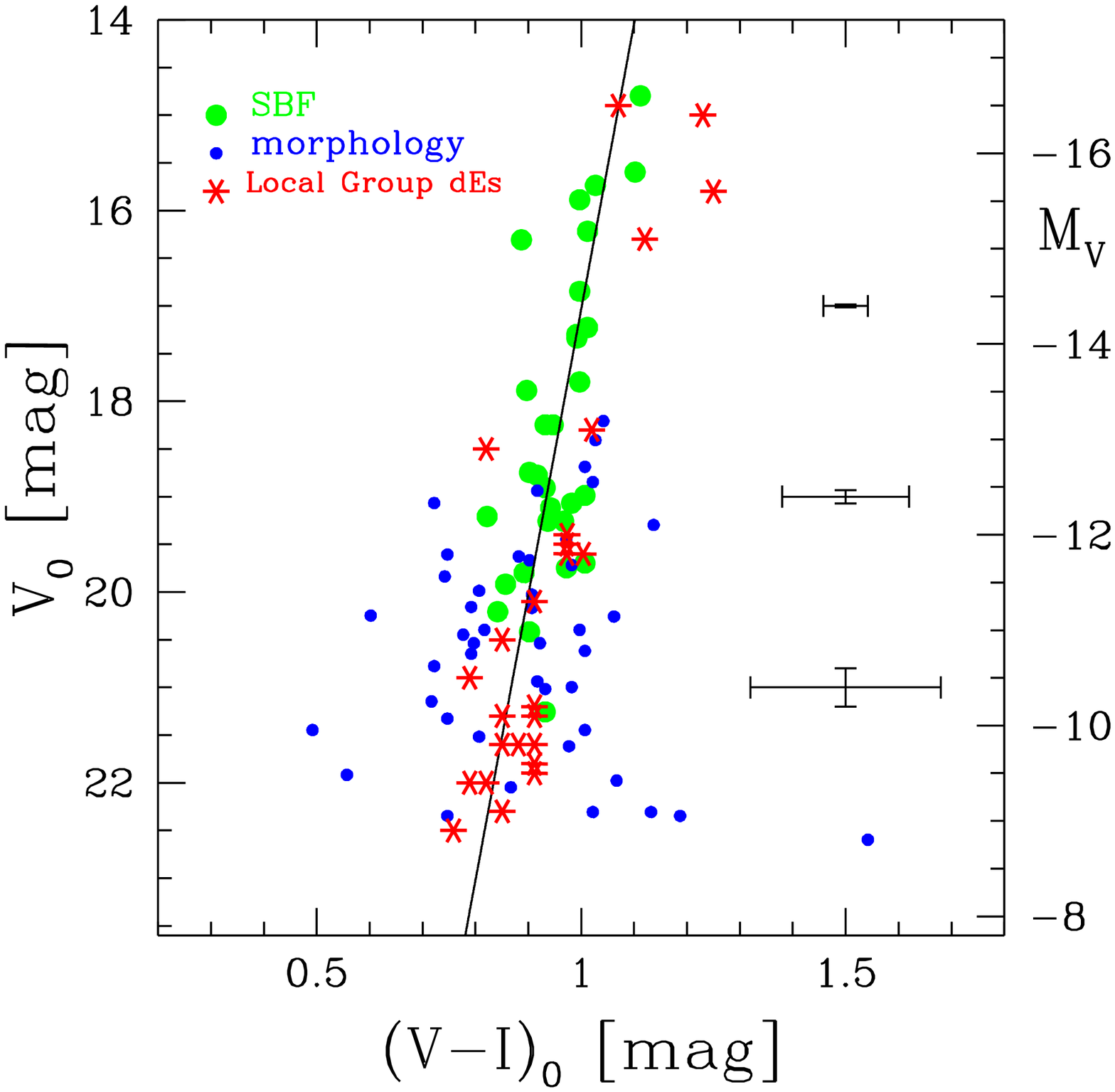} }
\resizebox{6.5cm}{!}{\includegraphics{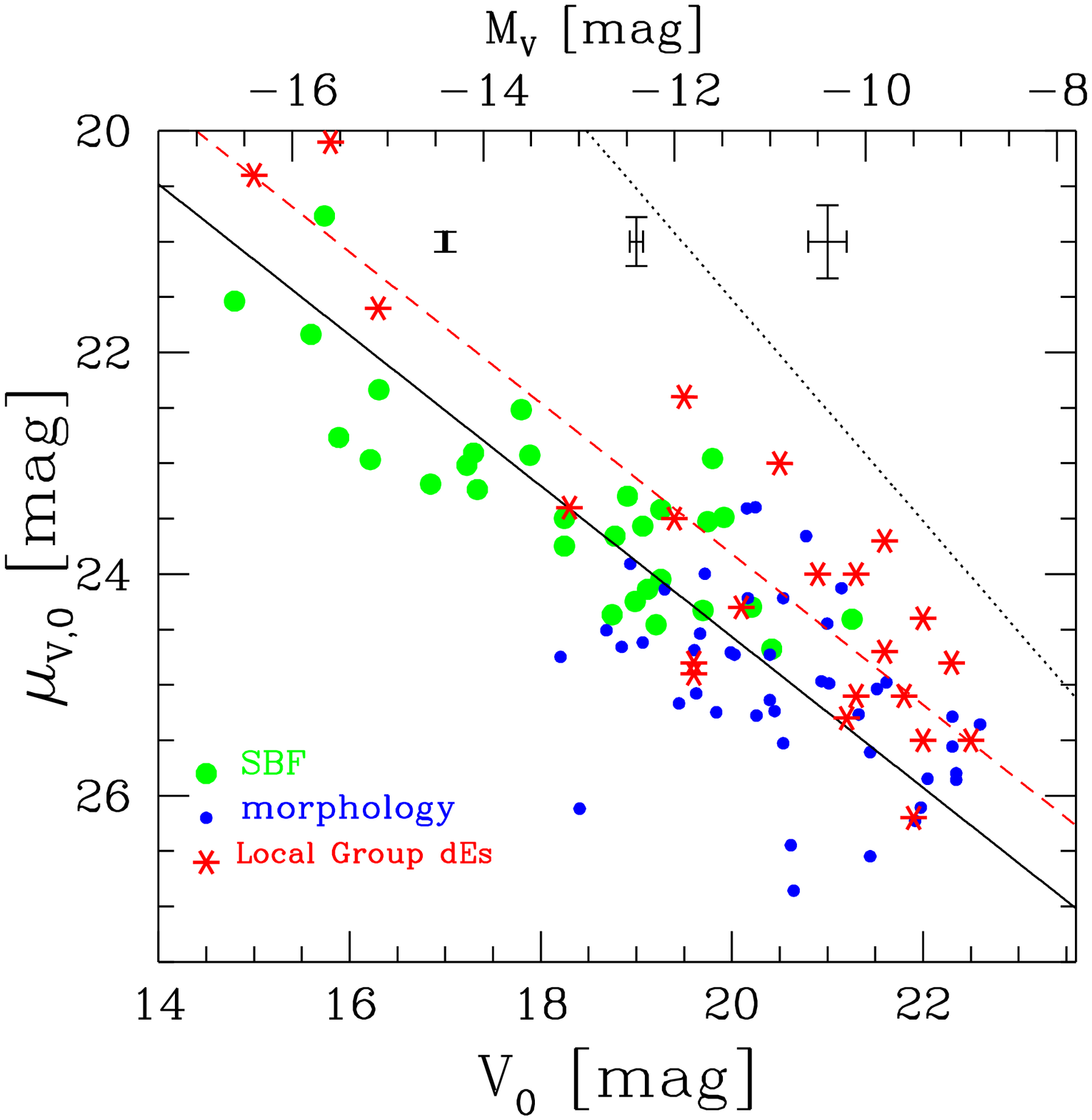} }

\caption[]{Colour-magnitude diagram (left) and surface brightness-magnitude
diagram (right) of Fornax cluster early type dwarf galaxies. Large filled
circles indicate dwarfs for which direct cluster membership via SBF measurement was obtained. Small filled circles indicate dwarfs for which the cluster membership assignment was via morphology. Asterisks indicate Local Group Dwarfs (Grebel {\it et~al.} 2003). In both plots, the solid lines are fits to the Fornax data points. In the right plot, the dotted line indicates the resolution limit of our data. The dashed line is a fit to the Local Group data points.}
\label{fig2}
\end{figure}
\begin{figure}
\centering
\resizebox{10cm}{!}{\includegraphics{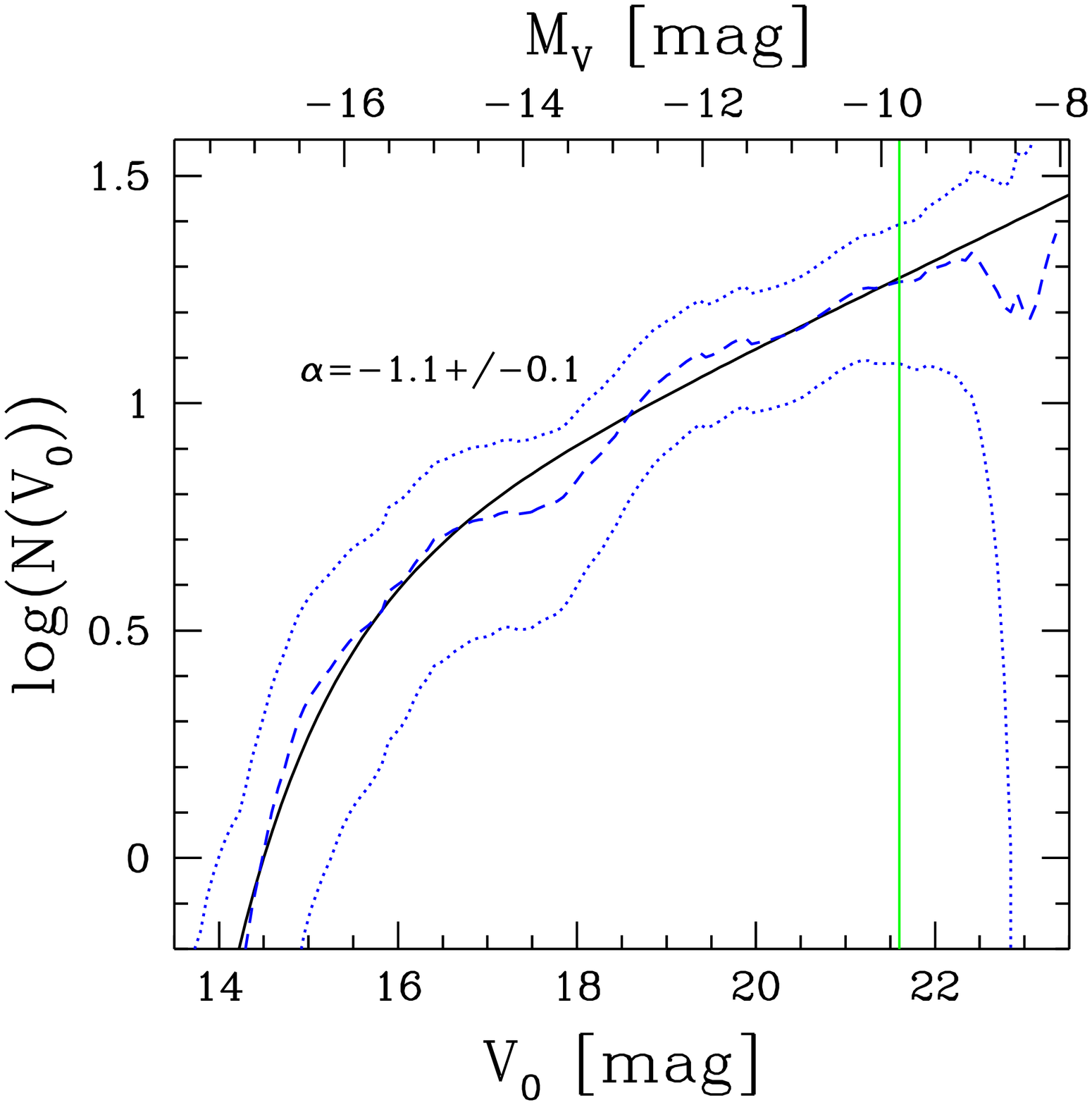} }

\caption[]{Incompleteness corrected luminosity function of the Fornax early-type dwarf sample as shown in Fig.~\ref{fig2}. The 50\% photometric completeness limit is
indicated by the vertical (green) line at $M_V \sim -$10 mag. The dashed line
is a binning-independent representation of the luminosity function, framed by its 1$\sigma$ uncertainty range (dotted lines). The solid line is a fit of a Schechter function, which yields a rather shallow faint end slope of $\alpha \sim -1.1$.}
\label{fig3}
\end{figure}

\section{Results}

We confirm cluster membership for 28 candidate dEs in the range
$-16.6<M_V<-10.1$ mag by means of SBF measurement. Of 51 further
candidate dEs in the range $-13.2<M_V<-8.6$ mag, 2/3 are confirmed as
probable cluster members by morphological re-assessment, while 1/3 are
re-classified as probable background objects. Finally, we find 12 new
dE candidates in the range $-12.3<M_V<-8.8$ mag.

In Fig.~\ref{fig2} we show the colour-magnitude and surface
brightness-magnitude relation of the resulting fiducial sample of
Fornax dwarf elliptical galaxies. The colours of the Fornax dwarfs are
very similar to their Local Group counterparts. Only the brightest
Local Group dwarfs appear slightly redder (more metal-rich). The sizes
of Fornax dwarfs are systematically larger than those in the Local
Group by about 40\%. This higher compactness in the Local Group may
have favoured self-enrichment more than in Fornax, possibly explaining
the metallicity offset of the brightest Local Group dwarfs.

Fig.~\ref{fig3} shows the incompleteness corrected GLF of the sample.
The derived faint end slope is $\alpha=-1.1 \pm 0.1$. Our
results thus confirm a very shallow faint end slope for the Fornax
dwarf galaxy luminosity function, in agreement with early estimates in
the reference study of Ferguson \& Sandage~(\cite{Fergus88}). 
We are therefore confident that this
value is robust and not very biased by systematic effects.
Morphological cluster membership assignment in Fornax apparently is
very reliable, provided that the image resolution is sufficient.

\begin{figure}
\centering
\resizebox{6.5cm}{!}{\includegraphics{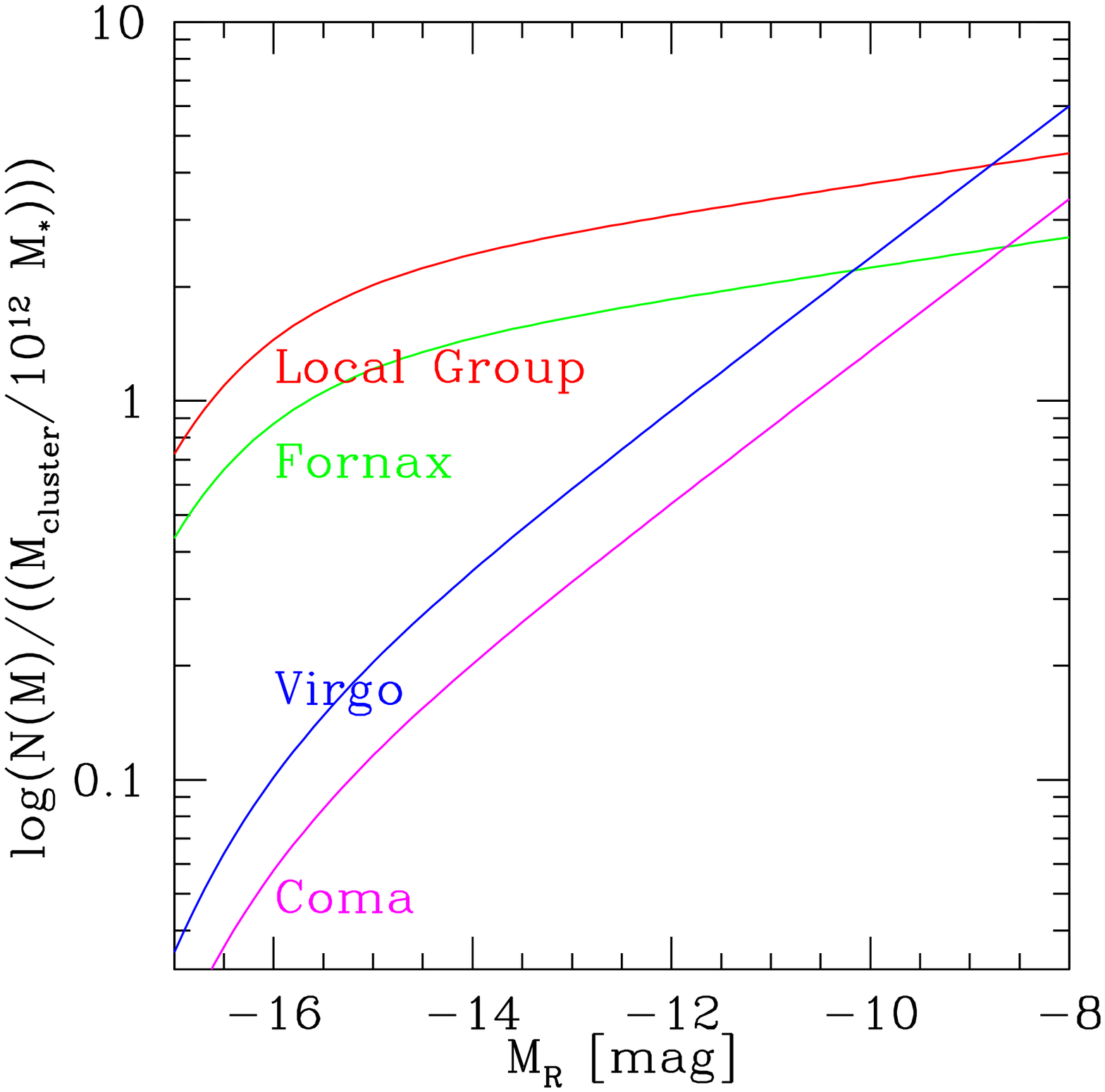} }
\resizebox{6.5cm}{!}{\includegraphics{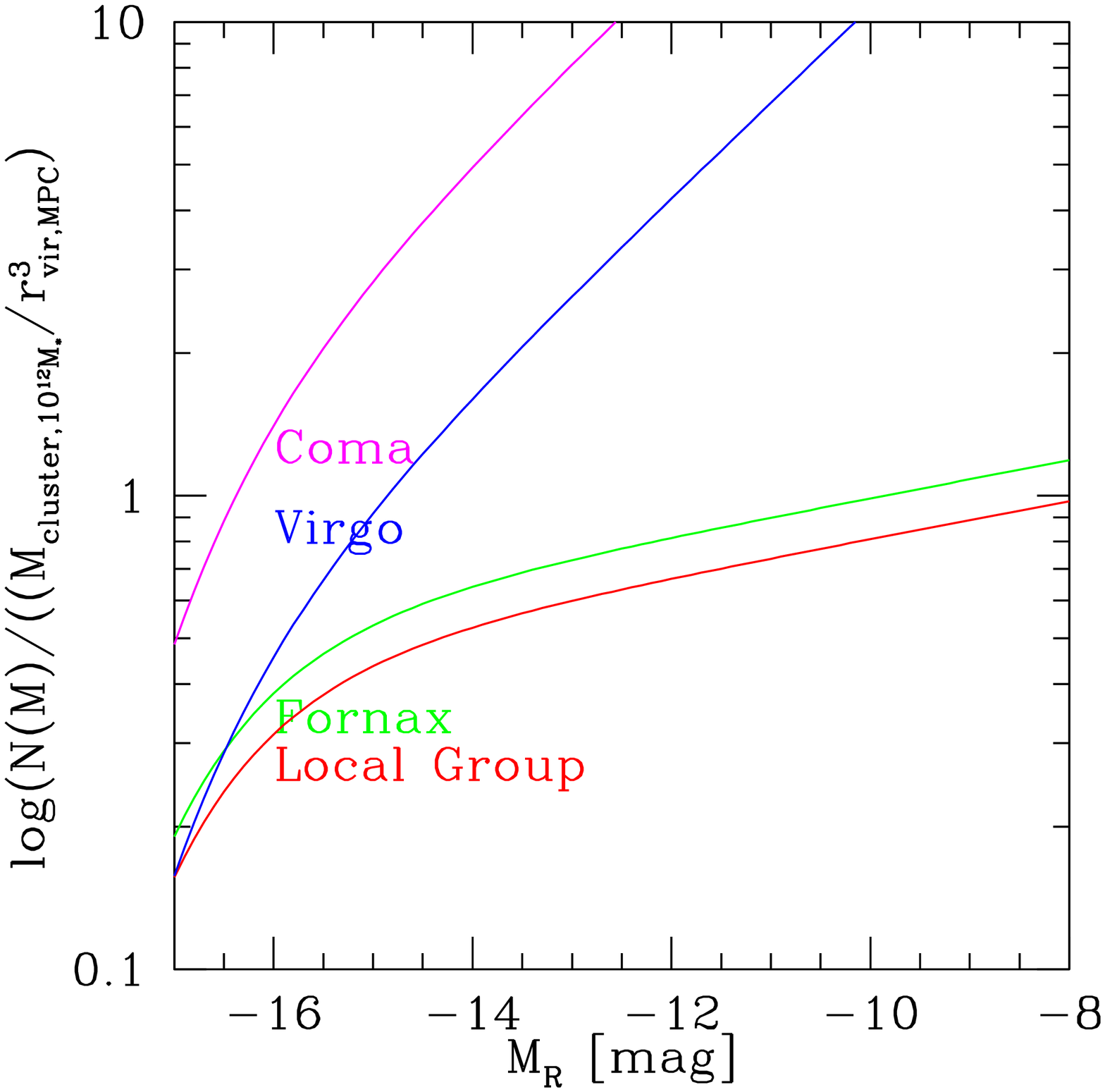} }

\caption[]{{\bf Left:}  Dwarf galaxy luminosity functions within the virial radii for four different environments (Local Group, Fornax, Virgo, Coma), normalised by the virial mass of each environment. {\bf Right:} Same luminosity functions as left, but normalised by the mass density within the virial radius. References: Drinkwater {\it et~al.}~(\cite{Drinkw01}), Trentham {\it et~al.}~(\cite{Trenth05}), Mamon {\it et~al.}~(\cite{Mamon04}), Sandage {\it et~al.}~(\cite{Sandag85}),  Ferguson \& Sandage~(\cite{Fergus88}), Mieske {\it et~al.}~(\cite{Mieske07}), Tonry {\it et~al.}~(\cite{Tonry00}), Mobasher {\it et~al.}~(\cite{Mobash02}) and references therein, Lokas \& Mamon~(\cite{Lokas04}), Sabatini {\it et~al.}~(\cite{Sabati03}).}
\label{fig4}
\end{figure}

\section{Discussion and conclusions}
It is well known that such a shallow faint end slope sharply
contradicts the much steeper value predicted for the mass function of
$\Lambda$CDM halos (e.g.Kauffman {\it et~al.}  \cite{Kauffm00}, Moore
{\it et al.}~\cite{Moore99}). Possible reasons for that discrepancy
include the accretion scenario (e.g. Hilker {\it
  et~al.}~\cite{Hilker99c}, C\^{o}t\'{e} {\it et~al.}~\cite{Cote98}),
where dwarf galaxies that fall into the cluster centres are tidally
disrupted, hence contributing to forming the extended cD halos of the
most massive cluster galaxies like NGC 1399 in Fornax. The presence of
ultra-compact dwarf galaxies (UCDs) in the central Fornax cluster
(Hilker {\it et~al.}~\cite{Hilker99b}, Drinkwater {\it et
al.}~\cite{Drinkw03}) may be a signpost of these tidal interactions
(Bekki et al.~\cite{Bekki03}, Mieske {\it et~al.}~\cite{Mieske06a}).

In Fig.~\ref{fig4} we compare the Fornax cluster dwarf GLF with galaxy
luminosity functions in other environments: the Local Group, the Virgo
cluster, and the Coma cluster. The absolute scale of all four
luminosity functions is from within the virial radius. In the left
plot, the luminosity functions are normalised by the dynamically
derived virial mass of each cluster. Interestingly, the number of
dwarfs per unit mass is highest in the lowest mass environment (Local
Group). It is lowest in the highest mass environment (Coma). This is
consistent with the finding that the M/L ratio of galaxy
groups/clusters increases with group/cluster mass (e.g. Eke {\it et
al.}~\cite{Eke06}). Another way of looking at this is that
a steep faint end slope does not imply a higher number of dwarfs relative
to the host cluster mass, rather the opposite. In the right
plot of Fig.~\ref{fig4}, we normalise the luminosity functions by mass
density of its host environments.  In this representation, environments
with a steeper faint end slope have produced more dwarf galaxies. 

\vspace{0.3cm}

We conclude that the SBF method is a very powerful tool to help
constrain the faint end of the galaxy luminosity function in nearby
galaxy clusters.  For the Fornax cluster, morphological cluster
memberships -- if performed at sufficient resolution -- are very
reliable.



\end{document}